# Bright entangled photon source without stringent crystal temperature and laser frequency stabilization

SANDEEP SINGH,[1,2,][*] VIMLESH KUMAR,[1] ANIRBAN GHOSH,[1] G.K. SAMANATA[1]

[1]*Photonic Sciences Laboratory, Physical Research Laboratory, Navarangpura, Ahmedabad, 380009 Gujarat, India*
[2]*Indian Institute of Technology-Gandhinagar, Ahmedabad 382424, Gujarat, India*
*\*Corresponding author: sandeep@prl.res.in*

**Entangled photon sources (EPS), the major building block for a variety of quantum communication protocols, are commonly developed by utilizing the spontaneous parametric down-conversion (SPDC) in $\chi^{(2)}$ nonlinear bulk optical materials. While high nonlinearity and long interaction length have established the superiority of the periodically poled crystals for EPSs, the phase-matching condition of such crystals is very sensitive to the fluctuation of the crystal temperature and the laser wavelength. As a result, deploying such sources outside the laboratory, for example, satellite-based applications, demands a stringent mass and power budget, thus enhancing the mission complexity and cost. We report a bright, stable entangled photon source with a relaxed requirement of crystal temperature and laser wavelength stabilization. Using a periodically poled KTP crystal inside a polarization Sagnac interferometer producing degenerate, type-0 phase-matched entangled photon pairs at 810 nm in an annular ring, we have transformed the SPDC ring into a "perfect" ring with the help of two common optical elements, axicon, and lens. Despite the variation of the SPDC ring size from Gaussian to annular of different diameters due to the change of crystal temperature over 7°C, and laser wavelength over 300 GHz, we observe the size of the "perfect" ring to be constant. The new EPS, having a spectral brightness as high as 22.58 ± 0.15 kHz/mW collected using single-mode fiber with a Bell's parameter, S = 2.64 ± 0.05, and quantum state fidelity of 0.95 ± 0.02, requires a crystal temperature stability of ±0.8°C, almost five times relaxation as compared to the previous EPS. The generic scheme can be used for non-collinear SPDC photons in all crystals to develop EPS at any wavelength and timescales for resource-constrained applications. © 2022 Optica Publishing Group**

---

The most prominent manifestation of non-locality proposed by quantum mechanics, commonly known as quantum entanglement, has found its implications in various domains, including quantum information, quantum communication, quantum metrology, and quantum computation [1, 2]. The central point of convergence of all these quantum technologies is to enhance information security by establishing a global-scale network of quantum internet [3] interconnected through different nodes and channels at various locations of the earth while accessing the advantage of quantum communication. Quantum communication, widely known as quantum key distribution (QKD) [4], is considered superior to its classical counterpart due to the encryption of the information through the use of the resourcefulness of photonic entanglement. While the most fundamental need for the demonstration of discrete variable QKD is the entangled photon source (EPS) with high state fidelity, the long-distance QKD network through optical fibers or free-space channels demands high brightness EPS.

On the other hand, the limited line of sight in terrestrial communication and the transmission losses of optical fibers, typically 0.142 dB/km [5] even for the state-of-the-art fibers, restricts the effective distance of communication to a range of about 100 km only. The further expansion of ground-based quantum links in free space channels demands the need for quantum repeaters [5], which is still a technological challenge. A more feasible substitute for all these challenges is to establish satellite-based communications using entangled photons. The recent development in the field of satellite-based quantum communication has enabled free space QKD over 1120 km at a secret-key rate of 0.12 bits per second [4] and attracted different counties to such initiatives. The major hurdle for long-distance communication is the low bit rate due to various parameters influencing the channel losses. One of the straightforward approaches to deal with such limitations is to enhance the brightness of the EPS based on spontaneous parametric down-conversion (SPDC) in $\chi^{(2)}$ nonlinear optical crystals [2, 6].

Given the low parametric gain of the second-order nonlinear process, efforts have been made to enhance the brightness of the entangled photon source by replacing critically phase-matched crystals [7, 8] with quasi-phase-matched periodically poled crystals [9, 2]. Among all the nonlinear crystals, Potassium Titanyl Phosphate (PPKTP) crystals in type-II phase-matching geometry have been used widely for entangled photon sources [10] despite exploring the full benefits of long interaction length and the high nonlinear coefficient as achieved in type-0 phase-matching geometry. This is due to the practical limitations in separating and detecting collinear, co-polarized paired photons of the same wavelength individually. Recently we have devised a new experimental scheme to develop high brightness entangled photon source using PPKTP crystal in type-0 phase-matching geometry [2]. Lowering the crystal temperature below the degenerate phase-matching, we have transformed the collinear, degenerate photon pairs at diametrically opposite points into an annular ring.

However, the entangled photon sources based on periodically poled crystals are highly sensitive to the phase-matching condition. As a result, such sources demand high stability in both crystal temperature (~0.1° C) and pump wavelength. Although these stringent requirements are easily addressable in lab conditions, deploying such sources outside the lab, for example, satellite-based applications, demands a stringent mass and power budget [5], thus enhancing the mission cost. On the other hand, the laser diode commonly used as the pump laser for field deployment of the entangled photon source has intrinsic mode hopping and mode jitter [11]. Therefore, it is essential to develop entangled photon sources with relaxed stringent requirements. Recently, efforts have been made to develop an entangled photon source with low-temperature sensitivity [10] but restricted to a particular wavelength range using type-II phase matching of the PPKTP crystal. Here, we report on a generic approach to developing a stable, bright EPS with relaxed constrain in the stabilization of the crystal temperature, mode hopping, and jitter of the pump laser. Using a 20 mm long PPKTP crystal in type-0 phase-matching geometry inside a polarization Sagnac interferometer, we have transformed the SPDC ring into a "perfect" ring with the help of an axicon and Fourier transforming lens. The size and position of the "perfect" ring are insensitive to the change of the SPDC ring resulting from the crystal temperature fluctuation and laser mode hopping, making the source stable against external perturbations.

The schematic of the experimental setup is shown in Fig. 1. A continuous-wave, single-frequency (linewidth ~12 MHz) laser providing output power of 50 mW at 405 nm is used as a pump source. A combination of the $\lambda/2$ plate and the polarizing beam splitter (PBS) cube is used to control the pump power in the experiment. A second $\lambda/2$ plate is used to transform the horizontal polarization state of the pump into diagonal to maintain equal pump power of the clock-wise and counter-clockwise beams of the polarization Sagnac interferometer comprised of a dual-wavelength PBS (D-PBS) and two plane mirrors, M1-2. A 20-mm long and 2 x 1 mm$^2$ in aperture PPKTP crystal placed at the center of the Sagnac interferometer between mirrors, M1 and M2, is pumped from both sides using the plano-convex lens (L1) of focal length, $f$ = 150 mm. The PPKTP crystal has a single grating period, $\Lambda$ = 3.425 µm, and is housed in an oven with a temperature stability of ±0.1°C to produce degenerate down-converted photons at 810 nm owing to type-0 ($e \rightarrow e+e$), non-collinear, phase matching of pump photons at 405 nm. The dual-wavelength $\lambda/2$ plate (D-$\lambda/2$), at both 405 nm and 810 nm, transforms the polarization of both pump and the down-converted photons from horizontal to vertical and vice versa. The counter-propagating SPDC photons of the polarization Sagnac interferometer on recombination by the D-PBS result in a Bell's state [2]. After collimation by the lens, L1, and subsequent extraction by the dichroic mirror, DM, the down-converted photons in the annular intensity profile (see inset image) are transformed into an annular ring of fixed diameter using a combination of lens and axicon. The Fourier transform of the Bessel beam generated by the axicon placed after the lens [12, 13] is transformed into an annular ring at the back focal plane of the lens L2, of focal length, $f$=100 mm. The apex angle of the axicon is 178.4°C. The annular intensity profile of the down-converted photons is divided into two symmetric half rings (see the inset images) using the prism mirror, PM (gold-coated right-angle prism). The pair photons are collected by the fiber couplers, C1 and C2, placed at the focal plane of the lens, L2. The polarization states of the photons are studied using the combination of $\lambda/2$ plate and PBS. The interference filter, F, of spectral width 3.2 nm centered at 810 nm, is used to extract down-converted photons from the background. The single-photon counting modules, SPCM1-2, connected to the fiber couplers, C1 and C2, using SMF or multimode fibers, and the time-to-digital converter (TDC) are used to measure the single-photon and coincidence counts. As reported previously [12], the size of the annular intensity profile and corresponding photon density depends on the position of the axicon from the Fourier plane. However, placing the axicon close to the Fourier plane to increase the overall photon counts adds mechanical constraints to the detection and analyzer systems. In such a case, a 4-$f$ imaging system (not shown here) is used to collect and analyze the down-converted photons. The temporal coincidence window for all the measurements is 1.6 ns.

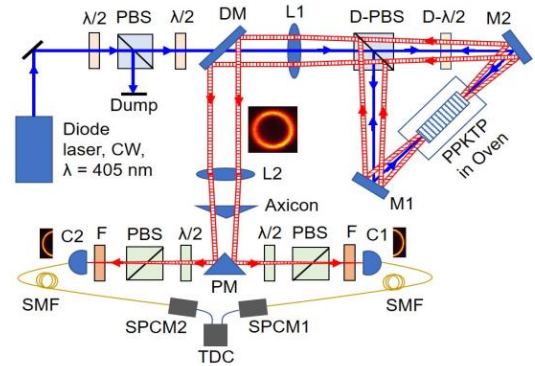

Fig.1: Schematic of the experimental setup. $\lambda/2$: half-wave plates; PBS: polarizing beam splitter cube; DM: dichroic mirror; L1-2: lenses; M1-2: mirrors; PPKTP: periodically poled KTP crystal; F: interference filter; PM: prism mirror, C1, 2: fiber couplers, SMF: single-mode fiber, SPCM1-2: single-photon counting modules; TDC: time to digital convertor. (Inset) ICCD image of the spatial profile of single photons.

To verify the concept of the SPDC perfect ring, we first measured the spatial distribution of the down-converted photons before the lens, L2, while varying the crystal temperature. The results are shown in Fig. 2. For this study, we have used the polarization state of the input laser to be either horizontal or vertical. As evident from the first row, (a-e), of Fig. 2, the diameter of the annular ring decreases with the crystal temperature from 21°C, resulting in collinear down-converted photons in the Gaussian spatial profile owing to the phase-matching conditions [14]. The variation in the cone angle or the diameter of the annular ring of the SPDC photons with temperature is the same as that of the previous report [2]. However, the shift in the temperature of the degenerate collinear SPDC photons from 36°C to 28°C can be attributed to the change in the laser wavelength. It is also evident from the first row of Fig. 2 that the collection optics (see green circle) set for optimum coupling of the diametrically opposite points of the SPDC ring no longer show optimum coupling for the SPDC rings at different crystal temperatures. On the other hand, as shown by the second row (f-j) of Fig. 2, the diameter of the perfect ring recorded at the Fourier plane of lens L2 is constant with the variation of the crystal temperature. The collection optics (see green circle) optimized for a crystal temperature collects diametrically opposite points of the SPDC ring for all crystal temperatures. To gain further insight, we have recorded the spatial profile of the SPDC ring and the perfect SPDC ring with the increase of crystal temperature from 21 °C with a step of 0.5 °C and measured the diameter and the width of the rings with results shown in Fig 2(k) and 2(l), respectively. As evident from Fig. 2(k), the radius of the SPDC ring (black dots) increases with crystal temperature, $\Delta T=T_o-T$, away from the degenerate collinear phase-matching temperature, $T_o=28$°C. However, the radius of the perfect SPDC ring (red dots) remains constant at ~ 2.73 mm. On the other hand, the width (FWHM) of the SPDC ring (black dots), as shown in Fig. 2(l), remains constant for all phase matching temperatures, but the width of the perfect SPDC ring (red dots) is constant (~200 µm) with temperature (up to 3°C) and increases to 380 µm at 7°C.

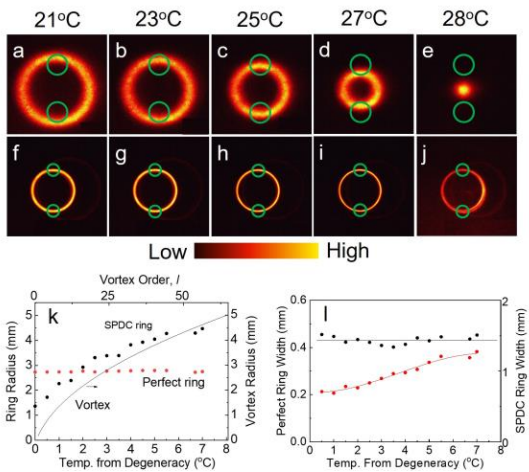

Fig. 2: Spatial distribution of the (a-e) SPDC photons and (f-j) perfect ring as a function of crystal temperature at a pump wavelength of 405.13 nm. (k) variation of radius of the SPDC ring (black dots) and the SPDC perfect ring (red dots) with the crystal temperature away from degeneracy at 28°C. The black line represents the variation of vortex beam radius with its order. (l) Variation of the width of the SPDC ring and perfect ring with the crystal temperature.

In literature, it is observed that the width of the perfect vortex increases with the order of the input vortex [15]. Therefore, to get a better perspective on the variation of the width of the SPDC perfect ring, we have calculated the vortex ring radius with the order using the Eq. (3) of Ref. [16] and fit (black line) to the experimental results (black dots) in Fig. 2(k). As evident from Fig. 2(k), the rate of increase of the SPDC ring radius with temperature is faster than that of the vortex ring radius with its order. The change in the SPDC ring radius for the crystal temperature 7° C away from the collinear phase-matching temperature (28° C) is equivalent to the change in the vortex ring for a vortex order increase of 50. Therefore, using the similar analogy of Ref. [15], we can confirm that the increase in the width of the SPDC perfect ring

(red dot of Fig (2l)) is due to the increase of the SPDC ring radius with crystal temperature. One can use the similar treatment of Ref. [15] to reduce the width of the SPDC perfect ring by increasing the radius of the perfect ring by simply moving the axicon close to the lens L2. However, such a proposition will reduce the photon number density collected by the fibers and the overall brightness of the paired photon source.

Similarly, we have studied the SPDC perfect ring with the change of the pump wavelength. Keeping the crystal temperature fixed at 25°C, we have pumped the crystal with three single-frequency lasers and measured the spatial profile of the SPDC ring before the lens, L1, and at its Fourier plane with the results shown in Fig. 3. The wavelength of the laser was measured using the high-resolution wave meter (High-finesse). As evident from the first row of Fig. 3, the spatial distribution of the SPDC photons changes from the collinear phase-matched Gaussian distribution to the non-collinear phase-matched annular ring of increasing radius for the increase of pump wavelength from 404.96 nm to 405.13 nm. Therefore, the collection optics (green circles) optimized for a particular pump wavelength (e.g., 405.01 nm) will miss the photons for the change of pump wavelength by 0.1 nm (~150 GHz) due to the mode hopping and mode jittering of the laser diodes. However, the radius of the SPDC perfect ring, as shown in the second row of Fig. 3, is constant despite the variation of the pump wavelength by ±150 GHz. Therefore, the collection optics (green circles) optimized for a fixed pump wavelength (say, 405.01 nm) remain optimized against the mode hop and wavelength jitter of the pump laser. Due to the unavailability of suitable laser diodes of different wavelengths in the lab, we could not test the current scheme for more wavelengths. However, the strong dependence of the phase-matching temperature on pump wavelength sets the ultimate limit for the allowed mode hopping wavelength range.

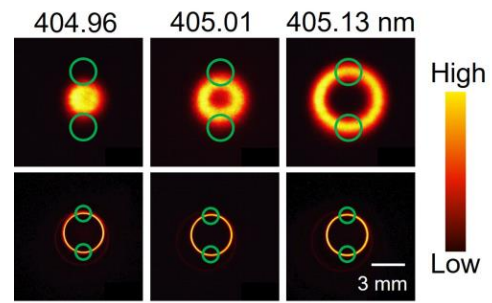

Fig. 3: Spatial distribution of the (top row) SPDC photons and (bottom row) perfect ring for the variation of pump wavelength at the crystal temperature of 25°C. The green circles represent collection optics.

To gain further insight, we have studied the coincidence counts of the photons collected from diametrically two opposite points on the SPDC ring and the perfect ring. Keeping the laser wavelength and the crystal temperature constant at 405.13 nm and 25 °C, we have measured the coincidence counts while varying the crystal temperature with the results shown in Fig. 4. As evident from Fig. 4(a), the coincidence counts (red dots) between the pair photons collected using the single-mode fiber from the SPDC ring optimized for crystal temperature decreases for the crystal temperature away from 25°C with a temperature bandwidth (FWHM) of 1° C. However, the coincidence counts for the photons collected using single-mode (black dots) and multimode (blue dots) from the

SPDC perfect ring have a temperature bandwidth of 3.2°C and 6.5°C, resulting in a 3.2 and 6.5 times enhancement in the temperature bandwidth. Further analysis reveals that a 10% peak-to-peak fluctuation in the coincidence counts requires the crystal temperature stabilization within ±0.15°C for the SPDC ring. However, using the SPDC perfect ring relaxes the temperature stability requirement of ±0.8°C, and ±1.25°C for photon collection using single-mode and multimode fibers, respectively.

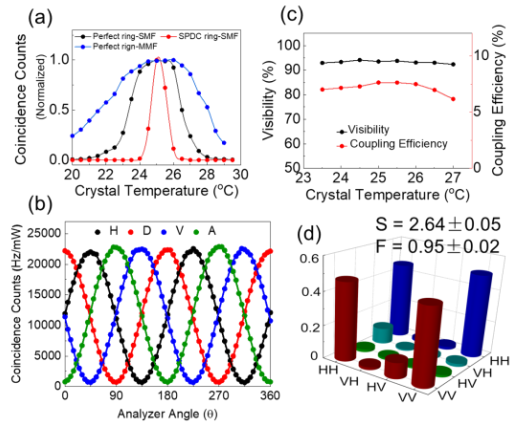

Fig. 4: (a) Variation of coincidence counts with crystal temperature. (b) Quantum interference of the entangled photons projected at different polarization states, H, D, V, and A. (c) Variation of quantum interference visibility and coupling efficiency with crystal temperature. (d) Graphical representation of the absolute values of the density matrix of the entangled photon states. Lines are the guide to the eyes.

Confirming the performance of the SPDC source of new architecture requiring relaxed crystal temperature control and laser mode hopping, we have studied the polarization entanglement using the standard coincidence measurement technique. Keeping the polarization of the pump laser diagonal, we have rotated the polarization state of one of the pair photons using the analyzer and measured the coincidence with its partner photon collected from the diametrically opposite point of the SPDC perfect ring at a fixed polarization state, H (horizontal), D (diagonal), V (vertical), and A (anti-diagonal). The results are shown in Fig. 4(b). As evident from Fig. 4(b), we observe a typical quantum interference of the polarized entangled photons for H (black dots), V (blue dots), D (red dots), and A (green dots) projections at the crystal temperature of 25°C. The solid lines are the best fit for the experimental data. The number of detected entangled paired photons have maximum and minimum values of ~2.2 x $10^4$ Hz/mW and 7 x $10^2$ Hz/mW for all polarization projections. The raw fringe visibility for polarization correlation is 94.2±0.06% in H/V bases and 93.8±0.06% in D/A bases confirming the non-local behavior of quantum entanglement. We also have calculated the Bell's parameter to be S = 2.64±0.03, corresponding to a violation of 21 standard deviations, confirming the generation of high-quality entangled photons. Further, we have measured the quantum interference visibility and the coupling efficiency (ratio of coincidence counts to the singles counts) of the source at different crystal temperatures with the results shown in Fig. 4(c). As evident from Fig. 4(c), the interference visibility and the coupling efficiency are constant at ~93% and 7%, respectively, for almost the entire temperature range, however, the coupling efficiency decreases with crystal temperature towards the collinear phase-matching (see Fig. 2). From this study, it is evident that the brightness of the entangled photons can further be increased by improving the coupling efficiency. Using the linear tomographic technique [17], we constructed the density matrix to find the degree of entanglement of the photon states with the results shown in Fig. 4(d). From the graphical representation of the absolute values of the density matrix of the generated states of Fig. 4(d), we determine the state to be, $|\psi\rangle = 1/\sqrt{2}(|HH\rangle - |VV\rangle)$, with a state fidelity of 0.95. As explained previously [2], we can transform the generated state to another Bell's state by adjusting the position of the crystal [18] in the Sagnac interferometer.

In conclusion, we have demonstrated a novel EPS suitable for resource-restricted applications. Pumping a 20-mm long PPKTP crystal placed inside a polarization Sagnac interferometer, using a cw diode laser at 405 nm, we have generated non-collinear, degenerate entangled photons at 810 nm over an annular ring. Using two simple optical elements, lens, and axicon, commonly available in an optics laboratory, we have transformed the annular ring into a "perfect" ring insensitive to the variation of the crystal temperature and laser wavelength. The generic experimental scheme, useful for the development of entangled photons at any desired wavelength and time scale, relaxes the stringent temperature stabilization of the crystal by five times and mode hopping and wavelength jitter by 300 GHz. Such demonstration opens up a new class of EPS for reliable deployment in the field demonstration of quantum optics experiments.

**Disclosures**. The authors declare no conflicts of interest.

**Data availability:** Data can be available under reasonable request.